\documentstyle[11pt,paspconf,epsf]{article}

\def\ltwid{\mathrel{\raise.3ex\hbox{$<$\kern-.75em\lower1ex\hbox{$\sim$}}}}
\def\gtwid{\mathrel{\raise.3ex\hbox{$>$\kern-.75em\lower1ex\hbox{$\sim$}}}}

\begin{document}

\title{Constraints on Disks Models of The Big Blue Bump from UV/Optical/IR
Observations}

\author{R. Antonucci}
\affil{Physics Department, University of California,
    Santa Barbara, CA 93106}

% Notice that some of these authors have alternate affiliations, which
% are identified by the \altaffilmark after each name.  The actual alternate
% affiliation information is typeset in footnotes at the bottom of the
% first page, and the text itself is specified in \altaffiltext commands.
% There is a separate \altaffiltext for each alternate affiliation
% indicated above.

%\altaffiltext{1}{}

% The abstract is entered in a LaTeX "environment", designated with paired
% \begin{abstract} -- \end{abstract} commands.  Other environments are
% identified by the name in the curly braces.

\begin{abstract}
Optical/UV observations provide many constraints on accretion disk models
of AGN which aren't always appreciated by modelers of the X-ray emission
(or sometimes even of the optical/UV emission).  The spectral behavior
at the Ly edge, the polarization, the continuum slopes and breaks, and the
variability timescales and phasing all conflict with simple models and
strongly constrain the more Baroque ones.  Partial-covering absorbers
and microlensing data suggest that the radiation is {\it not} released simply
according to where the potential drop (modified by standard viscous transport) 
takes place.  On the other hand, the orientation-based unified model is
in accord with the K-$\alpha$ inclination distributions for the AGN spectral 
classes, basing the latter on the limited existing data and theoretical 
understanding.

\end{abstract}

% Keywords should be included, but they are not printed in the hardcopy.

\keywords{quasar, Seyfert galaxy, accretion disk, black hole, polarization}

% That's it for the front matter.  On to the main body of the paper.
% We'll only put in tutorial remarks at the beginning of each section
% so you can see entire sections together.

\section{Introduction}

This talk considers first the constraints arising from direct observations of 
Type 1 AGN, including
quasars, and in the following section those deriving from the
inclination-based Unified Model (described in Antonucci 1993).

\section{Constraints from Direct Observations of Type 1 AGN}   

Here I use a list of diagnostics from a 1986 talk (published as Antonucci
1988) to show some historical development (how little has changed!), also appending more recent
information.

The Big Blue Bump optical/UV component, energetically dominant in quasars, was
first identified with thermal
emission from an accretion disk by Shields (1978).  AGN and quasars were
subsequently fit to model disks by Malkan, Laor, and many others.  These disks
were quasistatic and thin, powered by internal viscous dissipation.  They
radiated locally as blackbodies as assumed by Shakura and Sunyaev
(1973);
luminosities and spectral turnovers provided diagnostics of accretion rates and
black hole masses.

% In this section, we see the use of the \subsection command to set off
% an independent subsection.  We only have one here; usually there would
% be several.
%
% We show the use of several of the displayed math environments described
% in the User Guide, and you get a healthy dose of mathematical typesetting
% examples.  Also, observe the use of the LaTeX \label command after the
% \subsection to give a symbolic tag to the subsection for cross-referencing
% in a \ref command.  LaTeX automatically numbers the sections, equations,
% tables, etc. as it goes, so in general you don't know what number something
% is going to have.  We'll refer to the "hairymath" section a little later.

\subsection{Lyman Edge} 
 Kolykhalov and Sunyaev (1984) refined the models to include opacity effects,
using low-gravity stellar atmosphere models for the spectra of each disk annulus. 
They predicted enormous Lyman edges in absorption, contrary to observation. I was
encouraged to pursue this diagnostic by the following quotation from that
paper: ``A very important feature of our spectra is a considerable decrease of
the radiation flux beyond the Lyman continuum limit........ Different variants were
calculated to get a smaller Lyman discontinuity, but all attempts failed."  

We designed a study specifically to detect the broadened edges expected from
disk atmospheres, and
presented the results in Antonucci, Kinney and Ford (1989).
No {\it partial, broadened, systemic} edges  were found in emission or absorption, with
noise from the Lyman $\alpha$ forest limiting constraints to 10-20\%.  Later
Koratkar et al. (1992) found a very few {\it candidate} edges from IUE data.  These
await a
key test of an atmospheric edge as opposed to one produced by an
intervening cold cloud: there must be no accompanying narrow Ly absorption line
series with $\sim 10 ^{17} {\rm cm^{-2}}$ column density.

Subsequently, Tytler and Davis (1993) and 
Zheng et al. (1998) produced composites of low -Z quasar spectra with HST, the latter
claiming to see a weak feature near the edge position, but the former finding
no sign of any discontinuity or slope change. Caution is suggested regarding
the Zheng et al. claim.   The individual
objects in general look very significantly different from the composites (which
may be unphysical). Also, errors in Galactic reddening and the necessary
neglect of
host reddening is a very serious problem at these short wavelengths.  For a
very clean and high SNR spectrum of a BBB in the edge region, see Appenzeller
et al. 1998 ApJ \underbar {500}, L9: there is no feature, and interestingly, 
there is a rolloff in $\nu {\rm F} \nu$ near the Ly $\alpha$ 
position in the BBB of 3C273, indicating an unexpectedly
low temperature for the UV radiation.  Cataclysmic variables, which are known
to have ADs, generally show Ly and Ba edge features in absorption when optically thick
(in outburst), and in emission features when optically thin (in quiescence). The
observations are in qualitative agreement with the models.  Representative
references include Long et al. 1994, Kuigge et al. 1997, and Wagner et al.   
 1998; this is quite different from the AGN case.

Recent modeling has shown that the edges can be reduced by positing a
fast-rotating hole (e.g. Laor 1992),
Comptonization (Czerny and Zbyszewska 1991), consideration of metal  opacities 
especially if large turbulence is assumed (Agol et al. 1995; Krolik 1998; Blaes
1998 pc), and probably other effects.  A geometrically flat Comptonizing corona 
would conflict
with polarization data (see below).  This could be remedied by
Faraday-depolarization with an equipartition magnetic field (Agol and Blaes
1995). In that case it would then be
necessary to postulate subsequent
scattering by material between the disk and the broad line region in order to
produce the $\ltwid$1\% polarization, parallel to the radio jets, discovered
by Stockman et al. (1979).
                                
\subsection {Polarization}
Early models assumed pure scattering atmospheres for AGN accretion disks, which would lead to
substantial polarization (mostly 1-10\%, depending on inclination) parallel to the
disk plane.  Instead, $\ltwid$1\% polarization is observed, parallel to the
radio jets when the latter are observable (Stockman, Angel and Miley 1979,
discussed in this context by Antonucci 1988).  This polarization is associated
with the BBB, not the nonthermal radio core (e.g. see the SEDs in Antonucci,
Barvainis and Alloin 1990), and generally arises interior to the BLR (data of
Miller and Goodrich, partially displayed in Antonucci 1988).   (By contrast
little disk polarization is expected or observed in (dense) CV disks).

Laor et al. (1989) made modeling assumptions leading to very dominant absorption opacity in the
rest optical, but considerable Thompson-scattering polarization just below the
Lyman edge in frequency.  The incorrect direction of polarization was not
addressed.
I once mentioned to Joe Miller that this and other papers may have predicted the wrong
sign of polarization, but that they claimed to get the magnitude right.  He replied that
``that doesn't help when you balance your check book."

 More detailed calculations (Agol et al. 1998 and references
therein, and Matt et al. 1993),
showed that a given contribution of absorption to the opacity produces a
disproportionate reduction in polarization, contrary to the assumption of Laor et al.
However, most disk models find absorption contributes less to
the
opacity than those authors did.  It is also known that under certain conditions
a slight polarization can be produced which is parallel to the disk (and radiojet 
axis); see e.g. Gnedin and Silantev 1978.  However, this relies
on a very steep source function gradient, e.g. at annuli radiating in the exponential
tail of the spectrum, and probably isn't sufficiently general to explain the
observations.

As noted above, a favored solution may be complete Faraday depolarization of the
disk with a bit of ad hoc downstream scattering parallel to the disk plane (O. Blaes 1998, this meeting).
That final scattering, occurring interior to the BLR according to various
spectropolarimetry data, would need in essentially every case to lead to 0-1\%
polarization parallel to the disk axis.

A real surprise is the finding that substantial polarization can appear at
$\ltwid 700\AA$ (Impey et al. 1995; Koratkar et al. 1995, 1998).  The case of PG1630+377 from
the second paper is almost unbelievable, with P rising to $\sim 20\%$ before
the data cut off.  We plan to test this result, from the HST FOS
spectropolarimeter, with a different instrument.  The polarized flux spectrum
looks something like a Ly edge in emission, scattered by gas with $kT\sim 0.1
mc^2$.  In fact the object was selected for the ``missing" Ly continuum
photons (edge seen in absorption), 
our fond hope being that this indicated a bare accretion disk.   I
learned at the meeting that A. Beloborodo and collaborators are checking on
this possibility.  If it works out I'll feel better about our data set:
Eddington once said that he never believed an observation until it was
confirmed by theory. More likely (Beloborodo, pc;  Blaes and Agol, pc) the steepness of the rise
of the polarized flux with frequency, and the exact frequency of its onset, are
not fitted correctly at least in axisymmetric models. 

\subsection {Spectral Turnover for Low and High Luminosity AGN}
Active nuclei range over several orders of magnitude in luminosity, so an
inverse correlation between luminosity and temperature among the population, as
predicted by AD theory, 
should be easily detectable. (There could in principle be a finely tuned compensating trend in
L/L$_{\rm Edd}$.)  However, Seyferts and quasars generally show similar soft
x-ray excesses, indicating no strong differences in temperature if this is the
tail of the BBB (e.g. Walter and Fink 1993).  Within the rest optical/near-UV, 
quasars are actually flatter than Seyferts, corresponding to higher rather than lower
temperatures.

Some luminous quasars, studied in the context of the
$He^+$ Gunn-Peterson effect, have been observed to $\ltwid 300\AA$, and their
strength at such short wavelengths certainly
contradicts the thin blackbody disk (e.g. Reimers et al. 1989);  Comptonization
may save the day (Siemiginowska and Dobrzycki 1990), wiping out the Lyman edge
in the process.  Of course, a plain Comptonizing atmosphere or corona needs a
fine-tuned geometrical shape to produce the very low parallel polarization and 
to avoid major modifications of the disk-like Fe K$-\alpha$ lines.  Alternatively the
scattering polarization can, in principle, be destroyed by Faraday rotation.

Incidentally, the Comptonization (or optically thin thermal source in {\it
that}
model)  needs to be very special in order to lead
to nearly {\it the same effective T$_{\rm max}$} for objects over a huge
range of luminosity (Walter and Fink 1993).The rather fixed position of the soft excess 
suggests an atomic process,
perhaps broadened emission lines (Czerny and Zycki 1994) or albedo changes
(Czerny and Dumont 1998 and earlier references therein).    
                                                            
\subsection {Variability}
 Alloin et al. (1985) pointed out that the in-phase rapid variability of the
recombination lines (and thus the ionizing continuum) and the optical continuum is
inconsistent with the quasistatic viscous models of the day.  This type of
constraint, from direct multiband monitoring, is known to apply to quasars as
well (e.g. Cutri et al. 1985). Recent monitoring campaigns have tightened the
limit on any interband lags (e.g. Krolik et al. 1991), showing that in the disk model,
the various annuli must communicate at nearly light speed.  But the basic
problem has been known for a long time.  (In CV disks, flares may be rapid but
they generally propagate in wavelength and fade over time or reasonable viscous
timescales.)  Perhaps the x-rays only drive a variable part of the optical/UV;
however, the variability amplitude observed in the UV is often of order unity
even over just a few years (e.g. Cutri et al. 1985).  These observations 
have lead to a major paradigm shift to a passive disk,
heated from above, ostensibly by x-rays.  Aside from the fact that such a
model may predict large and unseen Lyman emission edges (Sincell and Krolik
1997), they are energetically ruled out for many Seyferts and all quasars
because the BBB is far more powerful than the x-ray (e.g. Laor et al. 1997). 
The only way out for quasars would be to postulate that the x-ray spectra are 
very different from 
those of Seyferts, with most of the
power in the hard x-rays where the observational limits are poor.  The energetics
{\it would} work for some low luminosity sources, and for some more, though far from all, if the x-rays are
beamed downward onto the disk by e.g. Compton scattering.  The latter idea is like
rearranging the deck chairs to keep the Titanic afloat.  It greatly
overpredicts the Compton hump in the x-ray spectrum (Malzac et al. 1998).

A separate variability issue, both for viscous disks and externally illuminated
ones, is the behavior of optical/UV color and the apparent temperature of the soft x-ray
excess as the luminosity of an object varies.  As a rule the optical/UV colors get bluer as an object brightens,
qualitatively consistent with thermal models.  In fact (to paraphrase Jim
Ulvestad), {\it all} AGN obey this rule,  except the ones that don't. 
Examples of the latter are Fairall 9 (Clavel et al. 1989; Rodriguez-Pascual et al.
1997; Santos-Lleo et al. 1997) and NGC4593 (Santos-Lleo et al. 1995). For those
which do get bluer, it remains to be shown that they do so in a way that
quantitatively matches expectations for optically thick thermal sources! 

The ``temperature" inferred from the soft x-ray  excess is generally very
constant as its flux undergoes large variations (e.g. Brandt 1998, this meeting). This is an important
puzzle for thermal models whose physical size is set by the region in which the
potential drop occurs, and hence which is constant over human time scales. The 
fixed position of the soft x-ray excess, among objects, and over time for
particular objects, suggests atomic emission or reflection to me (as advocated by Czerny and
Zycki 1994).
Atomic emission may be severely challenged by variability  
time scales (see the Brandt paper just cited), or observed spectra (modulo some ad
hoc Comptonization to hide emission lines), but the idea is worth further
analysis.
        
\subsection {Role of the IR Source}
                                                                           
Essentially all accretion disk models predict that F$_{\nu}$ increases with
$\nu$. (The few exceptions are cool and produce very few ionizing photons
directly, according to Blaes' talk at this meeting). Almost all quasars show
F$_{\nu}$ decreasing with $\nu$, after cutting out the small blue bump atomic
feature.  So how do the disk fits work?

The answer is that they generally rely on extrapolating an ``IR power law"
under the optical, subtracting it off, and fitting the difference.  Only
trouble is,  it
is now almost universally agreed that the IR component is thermal emission from
dust grains (blazars excepted).  Many good arguments lead to this conclusion, 
including e.g. low
frequency cutoff slopes, concommitant CO emission, very low polarization and
variability, the unified model and obvious thermal emission in Type 2s, and most 
crucially here, the universal inflection 
at $\sim 1\mu$  in the rest frames and the near IR reverberation mapping
(see e.g. Barvainis 1992 for details and references).  Dust emission cuts off
sharply at $\sim 1\mu$ because of sublimation.  (This is a good example of a  
universal frequency for a spectral feature indicating an atomic process.)  
Thus it is illegitimate to
subtract an ``IR power law" from the optical data.  This point is also relevant for
interpretation of  microlensing surface brightness constraints as described below.
                                                                                     
\subsection  {Opaque Partial Covering by Associated Absorbers}
Quasars have a well-studied subclass of narrow absorption lines among the so-called
``associated absorbers," generally (but not always) with nearly the systemic
redshift. These absorbers are ``macroscopic," arising on scales orders of
magnitudes larger than an accretion disk.  Yet some show unambiguous evidence
for opaque partial covering, that is, various doublet ratios show saturation,
so that $\tau \gg1$; at the same time they are {\it partial} rather than
black, typically absorbing only $\sim 50\%$ of the continuum.  Thus they are
thought to be opaque but to cover only part of the continuum sources.  
It is very doubtful that the light under the absorption lines can be
attributed to photons scattering around the clouds.  This does happen in some BAL
quasars and the scattered photons are highly polarized.  The very low continuum
polarizations of ordinary quasars, together with the moderate line depth,
proves generally that the residual photons which get through at the wavelengths of 
the absorptions
lines also have low polarization.  The
macroscopic cloud would have to be well centered on the AD in all cases, with
light scattering around it symmetrically, to exhibit such a low polarization. 

The implications for the AD model don't seem to be widely appreciated.  A
macroscopic absorber would need to have an atmosphere or edge (i.e. a region of intermediate optical
depth) small compared with the continuum source size,  and  cleanly cutting the
continuum source in two!  This is impossibly unlikely for a large foreground
cloud if all the radiation comes from $\sim R_g$ scales.  The near constancy
with time of
the absorption lines could also require that the cloud have virtually no
transverse velocity.

Less crazy, and interesting from the point of view of the nature of the
absorbers, is an array of tiny ($\sim R_g$) clouds with sharp edges in a stationary
state (as remarked by J. Krolik).  Here it might still be challenging to ensure
that a negligible fraction of their cross-sections have intermediate optical
depths.  This is necessary to preserve the saturated line ratios. 

This is one of several arguments that a significant part of the power may
escape the zone where the potential drop occurs without being converted into 
radiation.  The
case
of SS433 certainly shows that nature doesn't abhor this situation.
 
\subsection   {Microlensing Constraints}
This is an argument that a considerable fraction of the light of some quasars must
originate on a scale {\it too small} for an AD.  The argument is that the
rapid microlensing of the Huchra lens (aka Einstein Cross), Q2237+0305, sets a
lower limit to the surface brightness {\it greater than that of even a blackbody of
the same color temperature} (Rauch and Blandford 1991).Thus the thermodynamic 
emissivity of the disk model must be $\gtwid 10$!  The models apportioned
luminosity with radius in such a way as to reproduce the spectrum, thus
implicitly including energy transport (and thermalization) before radiation in a generic way.

One can relax the assumption that the model must fit the spectrum, and  then
successfully 
fit the variations at the observed wavelength (Jaroszynski, Wambgamss and
Pacynski 1992). I think the basic effect is that the disk temperature can be
higher than the observed color temperature.  The Rauch 
and Blandford paper was also criticized by e.g. Czerny et al. (1994) who were
able to fit the flux monitoring as well as the spectral data.  Their disk had
extra red flux from some somewhat ad hoc irradiation. This reduced the disk   
flux, and more importantly, increases the color temperature and hence
surface brightness.  Their disk also benefited slightly, through boundary
conditions, from allowing $L/L_{Edd}$ = 0.5 for a thin disk rather than 0.3 as in Rauch and
Blandford.  Finally, Czerny et al. implicitly allowed the fastest observed flux
variation to be atypically rapid (though not very rare), unlike Rauch and Blandford.

\section  {Indirect Constraints Based on Unified Models}
                    
\subsection {Expected Orientation of ADs of Type 1 Objects}
 Spectropolarimetric and other evidence indicates that the nuclear featureless
continuum sources in broad line regions are surrounded by $\gtwid$pc-scale
opaque dusty tori oriented perpendicular to the nuclear radio axes.  If the ADs
on much smaller scales are oriented similarly, than we observe them generally
at inclinations $\ltwid 45^{\circ}$. Thus we needn't be embarrassed if a sample
of such disks are all at relatively small inclinations, when using  diagnostics
such as the Fe K $\alpha$ line profiles.
 
\subsection { Orientation of ADs in Type 2 Objects}
It is well known that the nuclear jets and tori are not
preferentially oriented along the host spiral galaxies' axes and planes
(Ulvestad and Wilson 1984; Schmitt et al. 1997).  Thus if torus orientation
were the sole determinant of spectral classification; the ratio of the numbers
of the two types would be independent of host orientation.  That is not the
case however, with Seyfert 1's showing a particular aversion to edge-on hosts
(Keel 1980).  Thus, if tori are ubiquitous, some objects with face-on tori are 
still classified as type
2, because dust in the host galaxy plane obscures the nuclear featureless
continuum and broad line region (Lawrence and Elvis 1982).

A sample of Seyfert 2's, while probably primarily objects with highly
inclined tori (and by inference, ADs), will have significant contamination by
objects with face-on tori.  For certain studies these are imposters which can
spoil a sample; they generally have smaller x-ray columns than ``true" Seyfert 2's.
(They also would not be expected to show ionization cones.)
Turner et al. (1998) concluded, based on K$\alpha$ profiles of Compton-thin
Seyfert 2's, that they have face-on ADs just like the 1s. They may have been
 mislead                   
in part by neglecting this fact.  Their four ``Seyfert 2" objects with
K$\alpha$ lines seen directly, the keys to their conclusion, include NGC526A. 
It is really a Seyfert 1 (e.g. see the spectrum of Storchi-Bergmann et al. 1996)
in a nearly edge-on host.  I believe another problem with the conclusion of
Turner et al. is that there is no suitable (e.g. infrared-loud) numerically
sufficient parent population for the combined Seyferts if the latter are all
at~
$\ltwid 30^\circ$ inclinations as claimed.  See also Weaver and Reynolds (1998), 
who find much higher
inclinations than Turner et al. for the Seyfert 2 sample,  using the same data.

Why do we need to invoke tori at all for the ``imposter" Seyfert 2s, or for
that matter for all of the Seyfert 1s?  I think the most direct answer is that
the universal near-IR excesses (and the reverberation studies of those components) 
in Seyferts and quasars indicate substantial nuclear
covering factors of hot dust.

Another paper has appeared recently which claims to find evidence
restricting the generality of the Seyfert 1/2 unification, in this
case based on an HST imaging survey (Malkan, Gorjian and Tam 1998).
This result, and the little critique that comprises the rest of this
subsection, relate to ADs only indirectly, so may be skipped by the reader.

When testing a unification hypothesis, it's best to select the samples
by a property thought to be isotropic;  second best is to match by such
a property after the fact.  Neither was done in this paper; the targets
were just taken from the Veron-Cetty and Veron catalog of all known
objects.  Thus the results
can be suggestive at best.  The claims to be unbiased are unconvincing:
the radio and [O III] {\it fluxes} are claimed to be statistically
indistinguishable in their sample.  Aside from the fact that luminosities
would appear more relevant, statistical significance of a difference in
their sample is not the criterion for determining a bias - especially
since the ultimate conclusions are based on several-sigma differences at best,
which could be made insignificant by correction of even a small bias. Similarly the 
claim that the $12\mu$-selected sample (which is too small in itself to
be statistically significant) behaves similarly, does NOT ``avoid the
usual biases" as claimed, since in the unified models, the hot dust
    
emission is highly anisotropic. See e.g. Pier and Krolik (1993) ApJ 418, 673,
but the anisotropy is pretty much guaranteed in any model since the 
columns are generally so high that the dust is very opaque at that 
wavelength.  (The mere fact that hidden AGN have different infrared SEDs shows
the IR is anisotropic.)

The claim is also made by Malkan et al. that the Seyfert 1s in the
sample favor later Hubble types, and that this may mean some Seyfert BLRs
are obscured by material in the host plane rather than the nuclear torus.
As explained above, this has been known for at least 20 years.  See for
example the discussion in Antonucci 1993, Section 2.2.  Malkan et al. 
go on to say that nuclear dust lanes are more likely to
occur in Seyfert 2s than in 1s.  The difference is marginal statistically
and becomes insignificant when the samples are matched more closely
     
in redshift,
a key requirement for comparing small scale structures.  In fact it might
  
be harder to see such structures in Seyfert 1s even at the same redshift
since over half have saturated nuclei.  Their test of this effect seems to 
indicate that they ``missed only one sixth of [the dust lanes]" due to
simulated saturated point sources - but the effect would still seemingly be enough 
to reduce
the statistical significance of the result substantially. Similarly, limiting
the data to the 1s without detected point sources substantially reduces
the size of the effect, and renders it insignificant statistically.

The Malkan et al. paper has other puzzling statements, such as that the outer 
part of the torus is ``not expected to be much more than one or two orders
of magnitude larger [than the inner radius]," with no basis given.
However, the most bothersome error in the paper is the statement that
Seyfert 2 continuum polarizations are low, only ``up to 15\%", so that
the obscuring matter is geometrically thin in {\it most} cases.  
However, the polarization
    
papers have explained many times that the BLR polarization is generally
   
very high, with 16\% in NGC1068 being arguably the {\it minimum} observed.  In most
cases we just get an upper limit to BLR total flux ,and  thus just a lower limit to P.
The normal
Seyfert 1 seen in polarized flux is thus polarized at a high level 
in general, with a second {\it continuum} contribution  (sometimes referred to
as ``FC2") diluting the continuum
polarization.  It is now known that much of the dilution is from hot
stars (e.g. Heckman et al. 1997; Gonzalez-Delgado et al. 1998).

\subsection {Beamed X-rays in Radio Loud AGN}
Wozniak et al. (1998) review the x-ray spectral properties of broad line radio
galaxies, explaining that the contrast of the K$\alpha$ and Compton hump
features are lower than in Seyfert 1s.  Dilution by a beamed core component
could in principle be the cause.

In my opinion the unification by orientation of blazars with normal radio doubles
is most robustly inferred from a direct argument based on associated diffuse
radio emissions (Antonucci and Ulvestad 1985).  The excess x-ray  flux of
radio loud AGN relative to the radio quiet ones  or to  most lobe-dominant ones
closely tracks the beamed radio
core flux (e.g. Kembhavi 1993; Baker et al. 1995) so that
 in the Unified Model the radio core flux can be used to predict the level of
the beamed x-rays.   This provides another handle on possible dilution of the K
$\alpha$ equivalent width
 and Compton hump contrast.  

\subsection {D. Nonthermal AGN?}     
Finally tests of the unified model indicate a {\it possible} class of
radio galaxy without significant ``thermal" luminosity from accretion.  

The argument goes like this.  At the highest radio lobe luminosities, e.g. the distant
3C sources, the numbers of FR II radio galaxies and radio quasars are comparable,
and there is evidence that the projected linear sizes of the latter are
substantially smaller, in accord with the expected foreshortening (Barthel 1989). 
However, at lower luminosities (e.g. the nearby 3C sources), the FR II radio
galaxies greatly outnumber the quasars (e.g. Lawrence 1991; Kapahi 1990) and many
have small projected linear sizes compared with quasars (e.g. Singal 1993).    (I think that the
Molonglo survey data may be consistent with this general description also,
e.g. Kapahi et al. 1995.)

Many FR II radio
galaxies have been shown directly to harbor hidden quasars with
spectropolarimetry. 
However, the radio galaxies with small projected linear sizes, mostly ``optically dull"
(Laing 1994), might simply lack them.  Perhaps that would   
mean they are ``nonthermal AGN," e.g. powered by rotation,  via the
Blandford and Znajck (1977) mechanism. Alternatively, FR II radio galaxies may still have hidden quasars.
With the additional
supposition that torus opening angles increase with luminosity, and that the
individual object radio luminosities evolve in a reasonable way, Gopal-Krisha et al.
(1996) have shown that all the observed behavior described above is actually as
expected;
see also Punsly (1996) in
which detailed modeling leads to a similar conclusion.   
I am trying to observe them in the mid-IR to
look for the waste heat from the putative hidden quasars, and ISO may also
have something to say about this.  A serious x-ray survey on these (probably)
faint objects would be very valuable.
                                 
\section {Conclusion}                  
        
The arguments given are not airtight.  Perhaps I will write a paper attacking
this one, using an assumed name.  Nevertheless, they should be considered by
those who would invoke accretion disks for the Big Blue Bump in AGN: many AD
advocates have been notorious for failing to address counter arguments to the
model.

\acknowledgments

Participation in this meeting was supported by the organizers.  Research
described was funded largely by the NSF, most recently via grant number
AST-96177160.  Special thanks to Eric Agol, Richard Barvainis, Omer Blaes,
Roger Blandford, Bozena Czerny, Julian Krolik, Kevin Rauch, and Isaac Shlossman
for consultations.

% That's the end of the main body of the paper.  Now we will have some
% back matter.

% Now comes the reference list.  Since we typed out the citations ourselves,
% the reference list is enclosed in a "references" environment.  Each
% new reference begins with a \reference command which sets up the proper
% indentation.  Typography that may be required in the reference list by
% the editorial staff must be included by the author.
%
% Observe the "standard" order for bibliographic material: author name(s),
% publication year, journal name, volume, and page number for articles.
% Some journal names are available as macros; see the WGAS markup
% instructions for a listing of which ones have been "macro-ized".
% Note the use of curly braces to delimit the font changes: it is essential
% that this be done to limit the scope of the font declaration.
%
% There is no need to engage in any other typographic manipulation.

% That's all, folks.
%
% The technique of segregating major semantic components of the document
% within "environments" is a very good one, but you as an author have to
% come up with a way of making sure each \begin{whatzit} has a corresponding
% \end{whatzit}.  If you miss one, LaTeX will probably complain a great
% deal during the composition of the document.  Occasionally, you get away
% with it right up to the \end{document}, in which case, you will see
% "\begin{whatzit} ended by \end{document}".

\end{document}